\documentclass[prl,twocolumn,aps]{revtex4}

\usepackage{amsmath,amssymb}
\usepackage{latexsym}
\usepackage{graphicx}
\usepackage{color}

\begin{document}
\title{Nonlinear dressed states at the miscibility-immiscibility threshold}

\author{E. Nicklas$^1$, W. Muessel$^1$, H. Strobel$^1$,  
P.G.\ Kevrekidis$^{1,2}$\footnote{nlds@matterwave.de}, M.K. Oberthaler$^1$}

\affiliation{%
$^1$ Kirchhoff-Institut f{\"u}r Physik,  Universit{\"a}t Heidelberg, Im Neuenheimer Feld 227, 69120 Heidelberg, Germany
\\
$^2$ Department of Mathematics and Statistics, University of Massachusetts, Amherst MA 01003-4515, USA}

\begin{abstract}
\label{ssec:abs}

The dynamical evolution of spatial patterns in a complex system can reveal the underlying structure and stability of stationary states. As a model system we employ a two-component rubidium Bose-Einstein condensate at  
the transition from miscible to immiscible with the additional control of linear interconversion. 
Excellent agreement is found between the detailed experimental time
evolution and the corresponding numerical mean-field computations.
Analysing the dynamics of the system, we find clear indications of stationary states that we term nonlinear dressed states.  
A steady state bifurcation analysis reveals a smooth connection of these states with dark-bright soliton solutions of the integrable two-component Manakov model. 
\end{abstract}

\maketitle

%{\it Introduction.}  
Bose-Einstein condensates have been established over the past two decades as a prototypical testbed for exciting developments ranging from nonlinear dynamics and wave phenomena to superfluid features and quantum phase transitions ~\cite{Pethick-book,stringari,emergent,lcarr,prouk}. 
Especially two-component ultracold gases are ideal for the study of the connection of  topological solutions of integrable systems and their variants in the presence of different types of perturbations.

The properties of multi-component Bose Einstein condensates have been studied in numerous contexts.
In particular, early experimental efforts produced binary mixtures of two different hyperfine states of $^{23}$Na \cite{nake}  and of $^{87}$Rb~\cite{dsh}. 
The progressively improving experimental control has enabled detailed observations of phase separation phenomena and associated multi-component dynamics \cite{cornell,usdsh,wieman,hall2,hall3,tojo}. 
More recently, the mixing-demixing dynamics has been controlled both in pseudo-spinor (two-component)~\cite{flop} and  spinor systems~\cite{spielman} via external coupling fields.  
As a result, formation of domain walls has been observed. 
In these systems additional topological excitations such as dark-bright solitons do exist. 
These have been experimentally realized building on dynamical instabilities present in the regime of two counterflowing superfluids \cite{Engels}. 
The ability of phase imprinting offers a controlled path for the generation of individual such  topological states \cite{becker}.
All these observations are adequately captured by the mean-field description.
Thus, the well established integrable Manakov model \cite{ablowitz}, i.e. two nonlinearly interacting classical fields in one dimension at the miscibility-immiscibility threshold, forms a basis for  understanding the corresponding characteristics. 
This model is also examined in other physical systems such as  nonlinear polarization optics where multiple dark-bright and dark-dark soliton solutions can be systematically constructed~\cite{sheppard}.

%****************************************************************************
%***********************          FIGURE              ********************************
%****************************************************************************
\begin{figure}[tbph]
\begin{center}
\includegraphics[width = 6.5cm]{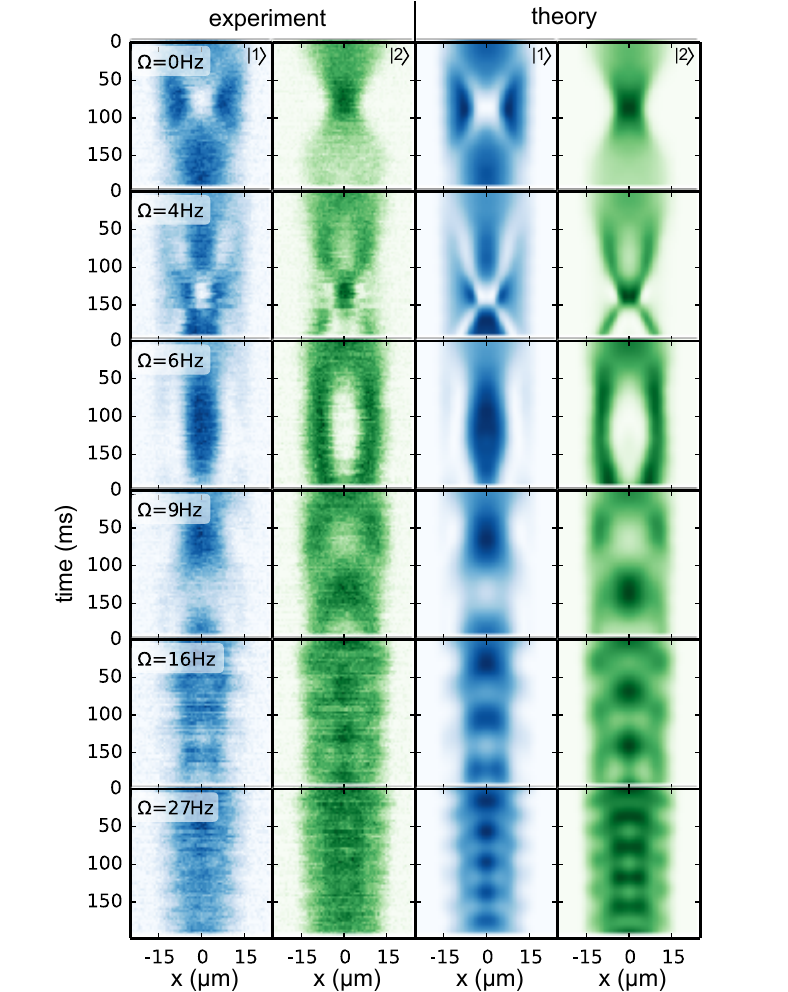}

\end{center}
\vspace{-0.4cm}
\caption{(Color online) Comparison of observed and numerically calculated time dynamics of an elongated two component condensate in the presence of a dressing field of different amplitudes $\Omega$. The asymmetry in the intra-species scattering lengths pushes component $|1\rangle$ to the wings of the trap for $\Omega = 0$. The trend reverses for 
$\Omega = 2 \pi \times 6 $\,Hz. As $\Omega$ increases further, the amplitude of the oscillatory dynamics decreases and in the large coupling limit the system is well approximated by stationary dressed states. 
}
\label{mofig1}
\end{figure}
%****************************************************************************
%****************************************************************************

Here, we study the nonlinear dynamics of a two-component Bose gas at the transition from  miscible to immiscible, arising through linear interconversion between the two components. 
In particular, we utilize a Rabi coupling between two hyperfine states of $^{87}$Rb and identify its significant impact on the dynamics as shown in Fig.~\ref{mofig1}. 
The comparison of experimental results with theoretical predictions shows excellent agreement. 
A more systematic analysis discussed below reveals that these observations can be understood as a consequence of the presence of stable nonlinear stationary states. 
These we will term ``nonlinear dressed states'' (NDS). 
Additionally, this new class of states is found to be interconnected as a function of the linear 
coupling strength via a series of Hamiltonian saddle-node, as well as Hopf bifurcations~\cite{royg}. 
A key observation is that the resulting rich bifurcation diagram connects the NDS with two previously studied limits. 
For vanishing linear coupling we recover the sequence of dark-bright solitonic states of the integrable Manakov model~\cite{sheppard} and in the limit of dominating interconversion we identify the known dressed states in the homogeneous miscible regime~\cite{search}.

In our experiment we initially prepare the gas in a product of single particle dressed states, i.e. an equal superposition of the two components, for given
Rabi coupling strength characterized by $\Omega$.
This is achieved by realizing a fast $\pi/2$ pulse with strong coupling and a subsequent phase-adjusted driving ($\phi=-\pi/2$) at the coupling strength of interest. It is important to note that with this procedure the higher excited states of the system are prepared. 
Fig.~\ref{mofig1} illustrates the comparison of the spatial dynamics for the two components, after a quench to different values of $\Omega$. 
The theoretical dynamics is based on the non-polynomial Schr{\"o}dinger equation (NPSE)~\cite{salas2}.
This confirms the quantitative relevance of utilizing the mean-field model in the regime under study as a suitable tool for predicting the dynamics. 

We observe that for $\Omega=0$, i.e. no linear coupling, component $|1\rangle$ is pushed to the edge of the cloud. This results from the fact that the repulsive interaction of component $|1\rangle$ is larger than for the other component $a_{11} > a_{22}$ ($[a_{11}, a_{22}, a_{12}] = [100.4, 95.0, 97.7] a_{\rm Bohr}$ ~\cite{usdsh,hall4}); here, $a_{xy}$ represents the scattering length between the $x,y$ components.
It is important to note that this is not due to demixing dynamics 
resulting from  an instability corresponding to $\Delta <1$~\footnote{Although in our case $\Delta < 1$, its proximity to
unity is such that the growth times of the most unstable modes are much longer than the time scale of our experiments.}  but has to be regarded as energetic separation of the two components. For the experiment described here, $\Delta  = a_{11} a_{22}/a_{12}^2 = 0.998(2) \approx 1$.
This trend is {\it reversed} as $\Omega$ is increased, where the more
strongly interacting component $|1\rangle$ is compressed 
during the dynamics initiated by the quench (see Fig.~\ref{mofig1} 
for $\Omega=2 \pi \times
6$\,Hz). This is a consequence of the finite size of the system and is well captured by the numerical calculations. 
For higher values of linear coupling we observe faster oscillatory dynamics which on average is reminiscent of the strongly dressed state regime~\cite{search} reported in the context of miscibility control by linear interconversion~\cite{flop}.

%{\it Methods.}
Before bringing these results into a more general context we briefly discuss the experimental and numerical methods used to monitor the system at hand. 
We create a Bose-Einstein condensate of  $5600$ $^{87}$Rb atoms in an elongated optical dipole trap with a longitudinal trap frequency $\omega_x = 2 \pi \times 23.4$\,Hz 
and a transverse confinement of $\omega_{\perp}= 2 \pi \times 490$\,Hz. The atoms are initially in state $|1\rangle =  |F = 1, m_F = -1 \rangle$ of the ground state hyperfine
manifold and can be linearly coupled to state $|2 \rangle = |F = 2, m_F = 1
\rangle$ via two-photon microwave and radio frequency radiation. The detuning 
from the intermediate $|2, 0\rangle$ level is  $-2 \pi \times 200\,$ kHz. 
A fast $\Omega \tau=\pi/2$ pulse with $\Omega=2 \pi \times 600$\,Hz creates a spatially homogeneous equal superposition of the two states. Within $5\, \mu$s  the phase of the coupling field is changed by  $-\pi/2$ and the amplitude is reduced to realize 
coupling strengths   $\Omega$ in the range of $2 \pi \times 0 \dots 60\,$ Hz. 
The magnetic field value of $B = 3.23\,$ G is chosen such that the differential 
Zeeman shift of states $|1\rangle$ and $|2 \rangle$ is equal to second order and the 
influence of magnetic field fluctuations is minimized.
The mean field shift due to the different intra-species scattering lengths
$a_{11} \neq a_{22}$ is compensated with a detuning of $\delta = -2 \pi \times
16$\,Hz.
The time evolution is obtained by repeating this  procedure and detecting the atomic clouds after different evolution times past the 
initial $\pi/2$ pulse using state-selective absorption imaging with a spatial resolution of $1.1$\, $\mu$m.

Our quantitative theoretical analysis is based on both a study of the system's time evolution and on the exploration of its stationary states and their 
Bogoliubov-de Gennes stability analysis. The time evolution of the linearly coupled atomic clouds is performed via the NPSE~\cite{salas2}. 
The simulation is initialized with the mean field ground state of $N = 5600$ atoms in state $|1 \rangle$ calculated via a Newton scheme. 
It subsequently  replicates the experimental procedure described above.
For the dynamical evolution in Figure \ref{mofig1}, we have also incorporated 
in the NPSE two- and three-body losses where the most important contribution comes from the spin relaxation loss of $F=2$.

%****************************************************************************
%***********************          FIGURE     2        ********************************
%****************************************************************************
\begin{figure}[tbph]
\begin{center}
\includegraphics[width = 8cm]{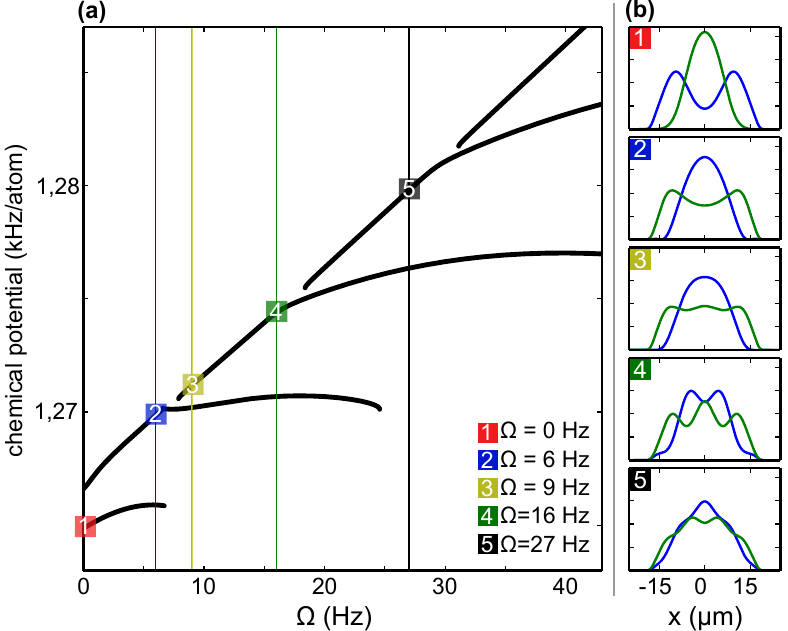}
\end{center}
\vspace{-0.4cm}
\caption{(Color online) Stationary states in the presence of a linear coupling field, exhibiting an intriguing cascade of branches.
The right columns show the theoretically obtained
density profiles for the two components $|1\rangle$ (blue) and $|2\rangle$ (green) for five characteristic values corresponding to 
the parameters of  Fig.~\ref{mofig1}. 
}
\label{mofig2}
\end{figure}
%****************************************************************************
%***********************          FIGURE   2          ********************************
%****************************************************************************

To shed light on the complex sequence of dynamical features observed for different values of $\Omega$, we proceed to  compute the stationary states of the coupled NPSE system
in Fig.~\ref{mofig2}. These are obtained by means of a fixed point (Newton) method,
capable of also capturing dynamically unstable states, a feature critical to our discussion below.
The stationary solutions are constructed by means of a small change to the parameter $\Omega$, using the previously converged stationary state as a seed. 
This parametric continuation approach reveals a sequence of branches which appear to be disconnected from each other, as can be seen in Fig.~\ref{mofig2}.
These form part of a progression whereby an increasing number of spatial
density modulations (and number of maxima) of each component is present in each higher branch; see Fig.~\ref{mofig2}(b). 
Notice, in particular, how component $|2 \rangle$ evolves from single hump for vanishing $\Omega$ (top panel of Fig.~\ref{mofig2}(b)) to double and multi-humped as $\Omega$ is increasing.

\begin{figure}[tbph]
\begin{center}
\includegraphics[width = 8cm]{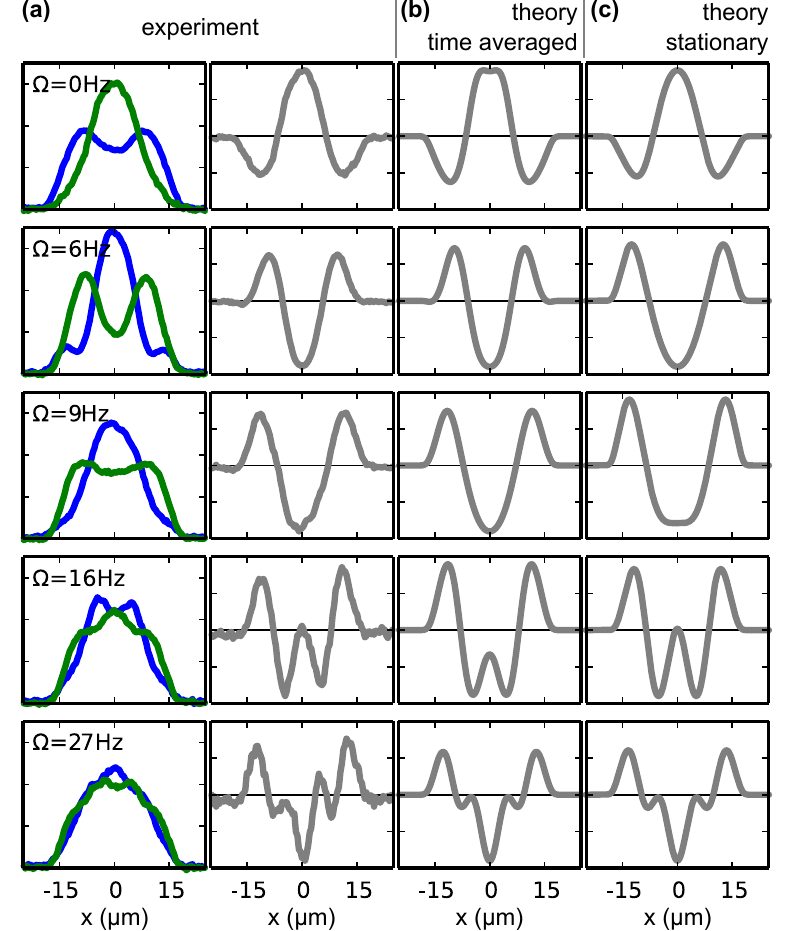}
\end{center}
\vspace{-0.4cm}
\caption{(Color online) Detailed comparison of the time-averaged density difference 
profiles with the stationary states. 
We extract the time averaged density profiles from the experimentally observed dynamics shown in Fig.~\ref{mofig1}. 
The same procedure is repeated for the numerical simulations and compared to the stationary states  from Fig.~\ref{mofig2} for the respective values of $\Omega$.
The agreement reveals that the time averaged profiles are accurately captured by the corresponding stationary nonlinear dressed states.}
\label{mofig3}
\end{figure}

We now show that these stationary states are intimately connected to the averaging of the experimental and numerical dynamical observations, as is illustrated in Fig.~\ref{mofig3}. 
For comparison, the population differences of the components are averaged over the respective period of the temporal evolution. 
Fig.~\ref{mofig3}(a) represents the experimental observations, panel (b) the direct numerical time dynamics (via the NPSE) while panel (c) depicts the corresponding stationary stable nonlinear dressed states; see also the discussion below. We find excellent agreement between time-averaged NPSE and the stationary NDS state predictions. Describing the observed dynamics as interference
of different stationary states at the same $\Omega$ with different
chemical potential, we expect that the averaging reveals the
strongest populated state. In our preparation procedure of a smooth
density distribution, this is given
by the stable nonlinear dressed states.

%****************************************************************************
%***********************          FIGURE   3          ********************************
%****************************************************************************
%****************************************************************************
%***********************          FIGURE   3          ********************************
%****************************************************************************

\begin{figure}[tbph]
\begin{center}
\includegraphics[width = 8cm]{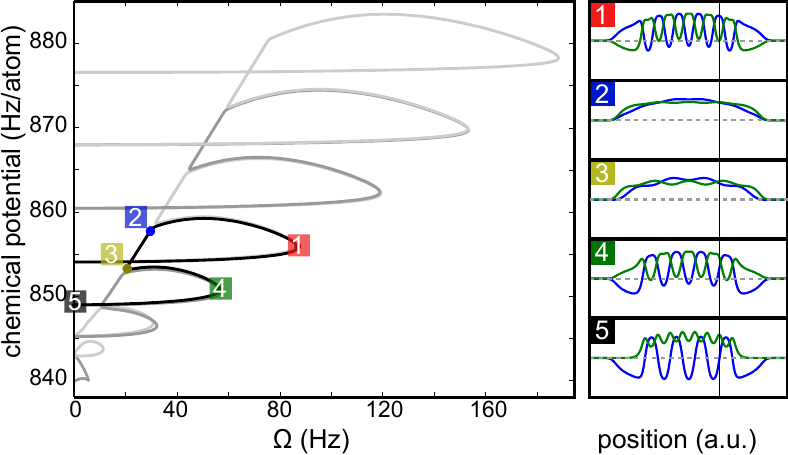}
\end{center}
\vspace{-0.4cm}
\caption{(Color online) The bifurcation loop structure for the stationary states.
The left graph shows the energy of the stationary states as a function of the linear coupling. The states are stable only along the linear parts, e.g. between markers 2 and 3, while the loops are unstable. 
The graphs at the right illustrate the spatial structure of the probability amplitudes of the corresponding stationary states. 
Following one specific loop we find that the nonlinear dressed states are smoothly connected to dark-bright soliton trains at $\Omega=0$ characterized by zero crossings of one component  and a corresponding amplitude maximum for the other component. 
The location of one of the dark-bright solitons in the state
identified by marker 5 is indicated by the vertical line.
Within each branch two additional topological excitations are added as the energy increases. 
}
\label{mofig4}
\end{figure}

In accordance with bifurcation theory, the depicted endpoints for our 
stationary solutions (see Fig.~\ref{mofig2}) cannot be isolated, but rather have to be continued.
To reveal this structure we utilize the method of pseudo-arclength 
continuation~\cite{doedel} enabling the following of the branches around these apparent endpoints. 
For this analysis we employed the one dimensional Gross-Pitaevskii equation
to facilitate the stability computations, yet the qualitative
observations reported 
below are unaffected by this.
Corresponding results including the spatial profile at selected points along one branch are shown in Fig.~\ref{mofig4}.
This illustrates that the nonlinear dressed states are smoothly connected to solutions of the dark-bright soliton type for vanishing $\Omega$ that are
known to exist in the context of the Manakov model~\cite{sheppard}. 
The depicted branch connects the state consisting of eight topological excitations to the one with ten; similar features arise for lower, as well as for higher branches (with corresponding soliton multiplicities). 
The symmetry of our initial guess selects the states with an even number of excitations.
The observed pattern in the experiment corresponds to the segment of the branch between the panels 2 and 3 in Fig.~\ref{mofig4}. 
The numerical Bogoliubov-de Gennes (linearization) stability analysis confirms that this segment is stable. 
The stability regime is delimited by a saddle node bifurcation at the lower corner (see e.g.  marker 3 in Fig.~\ref{mofig4}). 
This is characterized by a turning point of the branch connected with 
a zero crossing of an eigenfrequency in the Bogoliubov-de Gennes analysis. 
The upper limit of the stability segment (see e.g.  marker 4 in Fig.~\ref{mofig4}) is associated with a Hamiltonian Hopf bifurcation (see e.g.~\cite{royg} for a recent discussion), whereby quartets of eigenfrequencies emerge and destabilize the branch. 
Both unstable parts of the branch connect in the limit of $\Omega=0$ to a train of dark-bright solitons but with different even multiplicity. 
We emphasize that the stable stationary solutions naturally connect to the linear dressed states in the limit of large $\Omega$ \cite{search}. The amplitude of the spatial structure as well as its length scale decreases as $\Omega$ approaches this limit in accordance with the Bogoliubov-de Gennes analysis of the uniform state~\cite{tommasini}.

We note that for small linear couplings both the profile and the energy is strongly influenced by the finite size of the system. Especially the observed interchange of the components between $\Omega=2 \pi \times 4$\,Hz and 
$\Omega=2 \pi \times 6$\,Hz (see Fig.~\ref{mofig1}) can be attributed to the turning point of the lowest branch shown in Fig.~\ref{mofig4}.
The lower two branches in Fig.~\ref{mofig4} represent solutions whose spatial extent
(even with a single hump) induces competition with the spatial length allowed by the trap, hence their ``unusual'' shape.

%****************************************************************************
%***********************          FIGURE   4          ********************************
%****************************************************************************
%****************************************************************************
%***********************          FIGURE   4          ********************************
%****************************************************************************

%{\it Conclusions.}
In the present work, we have studied experimentally and theoretically the extension of the Manakov model by introducing linear coupling between the two components. We find stable stationary solutions for any value of the linear coupling which we term nonlinear dressed states. The theoretical identification is found to be in excellent agreement with our experimenal observations. Furthermore, we establish a connection to limiting solutions in the form of dark-bright soliton trains in the vanishing linear coupling limit and to linear dressed states in the large coupling limit. The associated branches reveal stable and unstable segments separated by saddle-node and Hamiltonian Hopf bifurcations. 

Our results reveal a previously unidentified connection between highly nonlinear stationary states and the weakly interacting regimes reached via a controlled perturbation. 
Identifying the nature of perturbations leading to such a smooth connection may provide critical insights into a physical system revealing universal characteristics for highly excited states. 
The translation of our findings to spin models, e.g. in our case the
mapping to a transverse Heisenberg model \cite{kasamatsu}, can prove fruitful in characterizing the high energy part of the excitation spectrum. 
This might have consequences on the recently discussed excited state quantum phase transitions \cite{caprio}. 
Moreover, it would be interesting to explore how such phenomenologies may extend to higher dimensional settings, potentially forming a multi-dimensional analog of nonlinear dressed states and a set of branches connecting 
different spatially modulated states.

The authors acknowledge experimental contributions by Ji\v{r}\'i Tomkovi\v{c}  
and fruitful discussions with Boris Malomed. This work was supported by the Heidelberg Center for Quantum Dynamics and the European Commission small or medium-scale focused research
project QIBEC (Quantum Interferometry with Bose-Einstein condensates, Contract Nr. 284584). W.M. acknowledges support by the Studienstiftung des deutschen Volkes. P.G.K. acknowledges support from the National Science Foundation under grant DMS-1312856, and from the US-AFOSR under grant FA9550-12-10332.


\begin{thebibliography}{99}

\bibitem{Pethick-book}C. J. Pethick and H. S. Smith,
\emph{Bose-Einstein Condensation in Dilute Gases}, Cambridge
University Press, Cambridge, 2002.

\bibitem{stringari} L.P. Pitaevskii and S. Stringari,
{\it Bose-Einstein Condensation}, Oxford University Press (Oxford, 2003).


\bibitem{emergent}
P.G. Kevrekidis, D.J. Frantzeskakis, and R. Carretero-Gonz{\'a}lez (eds.),
{\it Emergent nonlinear phenomena in Bose-Einstein condensates. Theory and experiment}
(Springer-Verlag, Berlin, 2008).

\bibitem{lcarr} L.D. Carr, {\it Understanding Quantum Phase Transitions},
(Taylor \& Francis, Boca Raton, 2010).

\bibitem{prouk}  N. Proukakis, S. Gardiner, M. Davis, M. Szymanska,
{\it Quantum gases: finite temperature and non-equilibrium dynamics},
(Imperial College Press, London, 2013).


\bibitem{nake} D.M.\ Stamper-Kurn, 
M. R. Andrews, A. P. Chikkatur, S. Inouye, H.-J. Miesner, J. Stenger, and 
W. Ketterle,
\newblock
Phys.\ Rev.\ Lett.\ \textbf{80}, 2027 (1998).



\bibitem{dsh} D.S.\ Hall, 
M.R. Matthews, J.R. Ensher, C.E. Wieman, and 
E.A. Cornell, \newblock Phys.\ Rev.\ Lett.\ 
\textbf{81}, 1539 (1998).




\bibitem{cornell} V. Schweikhard, I. Coddington, P. Engels, S. Tung
and E.A. Cornell, Phys. Rev. Lett. {\bf 93}, 210403 (2004).

\bibitem{usdsh} K.M. Mertes, J.W. Merrill, R. Carretero-Gonz{\'a}lez,
D.J. Frantzeskakis, P.G. Kevrekidis and D.S. Hall,
Phys. Rev. Lett. {\bf 99}, 190402 (2007).

\bibitem{wieman}  S.B. Papp, J.M. Pino and C.E. Wieman,
Phys. Rev. Lett. {\bf 101}, 040402 (2008).

\bibitem{hall2} R.P. Anderson, C. Ticknor, A.I. Sidorov,
B.V. Hall, Phys. Rev. A {\bf 80}, 023603 (2009).

\bibitem{hall3} M. Egorov, B. Opanchuk, P. Drummond, B. V. Hall, P. Hannaford, and A. I. Sidorov
Phys. Rev. A {\bf 87}, 053614 (2013).

\bibitem{tojo} S.Tojo, Y. Taguchi, Y. Masuyama, T. Hayashi,
H. Saito and T. Hirano, Phys. Rev. A {\bf 82}, 033609 (2010).

\bibitem{flop} E. Nicklas, H. Strobel, T. Zibold, C. Gross,
B.A. Malomed, P.G. Kevrekidis, M.K. Oberthaler,
Phys. Rev. Lett. {\bf 107}, 193001 (2011).

\bibitem{spielman} Y.-J. Lin, K. Jim{\'e}nez-Garc{\'i}a,
I.B. Spielman, Nature (London) {\bf 471}, 83 (2011).




\bibitem{Engels} C. Hamner, J.J. Chang, P. Engels, and M. A. Hoefer, 
%Generation of Dark-Bright Soliton Trains in Superfluid-Superfluid Counterflow,
 Phys. Rev. Lett. {\bf 106}, 065302 (2011);
M.A. Hoefer, J.J. Chang, C. Hamner, and P. Engels, 
%Dark-dark solitons and modulational instability in miscible two-component Bose-Einstein condensates, 
Phys. Rev. A {\bf 84}, 041605 (2011);
D. Yan, J.J. Chang, C. Hamner, P.G. Kevrekidis, P. Engels, V. Achilleos, D. J. Frantzeskakis, R. Carretero-Gonzalez, P. Schmelcher, 
%Multiple dark-bright solitons in atomic Bose-Einstein condensates, 
Phys. Rev. A {\bf 84}, 053630 (2011);
D. Yan, J.J. Chang, C. Hamner, M. Hoefer, P.G. Kevrekidis, P. Engels, V. Achilleos, D.J. Frantzeskakis and J. Cuevas, 
%Beating dark-dark solitons in Bose-Einstein condensates, 
J. Phys. B: At. Mol. Opt. Phys., {\bf 45} 115301 (2012).

\bibitem{becker} C. Becker, S. Stellmer, P. Soltan-Panahi, S. D¨orscher, M.
Baumert, E. M. Richter, J. Kronj{\"a}ger, K. Bongs, and K. Sengstock,
Nat. Phys. {\bf 4}, 496 (2008).
 

\bibitem{ablowitz} S. V. Manakov, Zh. Eksp. Teor. Fiz. {\bf 65}, 505 (1973) [Sov. Phys. JETP {\bf 38}, 248 (1973)]; V. E. Zakharov and S. V. Manakov, Zh. Eksp. Teor. Fiz. {\bf 71}, 203 (1976)
[Sov. Phys. JETP {\bf 42}, 842 (1976)]. For a more detailed 
discussion see e.g.
M.J. Ablowitz, B. Prinari and A.D. Trubatch,
{\it Discrete and Continuous Nonlinear Schr{\"o}dinger Systems},
Cambridge University Press (Cambridge, 2004).


\bibitem{sheppard} A.P. Sheppard, Yu.S. Kivshar,
Phys. Rev. E {\bf 55}, 4773 (1997).

%\bibitem{scattering_lengths}
%Yun Li, P. Treutlein, J. Reichel, and A. Sinatra. Spin squeezing in a bimodal condensate: spatial
%dynamics and particle losses. Eur. Phys. J. B, 68(3):365–381, 2009.

\bibitem{royg}  See e.g. for a recent discussion R.H. Goodman, 
J. Phys. A: Math. Theor. {\bf 44}, 425101 (2011) and 
references therein.


\bibitem{search} C.P. Search and P.R. Berman,
Phys. Rev. A {\bf 63}, 043612 (2001).

\bibitem{salas2} The one component version of the model was derived in
L. Salasnich, A. Parola, and L. Reatto, Phys. Rev. A {\bf 65}, 043614 (2002),
while the two-component version used herein was obtained in
L. Salasnich and B. A. Malomed, Phys. Rev. A {\bf 74}, 053610 (2006).

\bibitem{hall4} A. M. Kaufman, R.P. Anderson, T. M. Hanna, 
E. Tiesinga, P. S. Julienne, and D. S. Hall, Phys. Rev.
A {\bf 80}, 050701 (2009).

\bibitem{doedel} E.J. Doedel, {\it Lecture Notes on Numerical
Analysis of Nonlinear Equations}, available at:
http://cmvl.cs.concordia.ca/publications.html.

\bibitem{tommasini} P. Tommasini, E.J.V. de Passos, A.F.R. de Toledo Piza,
M.S. Hussein, E. Timmermans, Phys. Rev. A {\bf 67}, 023606 (2003).

%\bibitem{spielnew} S. De, D. L. Campbell, R. M. Price, A. Putra, B. M. Anderson, I. B. Spielman, arXiv:1211.3127.


\bibitem{kasamatsu} K. Kasamatsu, M. Tsubota, M. Ueda, Phys. Rev. A {\bf 71}, 043611 (2005).

\bibitem{caprio} M.A. Caprio, P. Cernar, and F. Iachello,
Ann. Phys. (N.Y.) {\bf 323}, 1106 (2008).



\end{thebibliography}
\end{document}